\documentclass{JINST}
\def\NIMA#1#2#3{{\rm Nucl.~Instr.~and~Meth.} {\bf{A#1}} (#2) #3}

\newcommand{\etal}{et al.}
\newcommand{\alfa}{ALFA}
\newcommand{\alfaA}{ALFA 2\_2\_64}
\newcommand{\alfaB}{ALFA 10\_2\_16}
\newcommand{\OD}{OD}

\title{Hadron beam test of a scintillating fibre tracker system for elastic scattering and luminosity measurement in ATLAS}
\author{
F.~Anghinolfi$^a$,
S.~Ask$^a$\thanks{Corresponding author.},
P.~Barrillon$^b$,
G.~Blanchot$^a$, 
S.~Blin$^b$,
A.~Braem$^a$,
C.~de La Taille$^b$,
B.~Di Girolamo$^a$,
I.~Efthymiopoulos$^a$,
J.~Faustino$^{c,e}$,
D.~Fournier$^b$,
S.~Franz$^a$,
P.~Grafstr\"om$^a$,
L.~Gurriana$^e$,
M.~Haguenauer$^f$,
V.~Hedberg$^g$,
M.~Heller$^b$, 
S.~Hoffmann$^h$, 
W.~Iwanski$^{a,i}$,
C.~Joram$^a$,
A.~Ko\v{c}n\'{a}r$^j$,
B.~Lavigne$^b$,
B.~Lundberg$^g$,
A.~Maio$^{d,e}$,    
M.J.P.~Maneira$^c$,  
A.~Mapelli$^a$,
C.~Marques$^e$,
U.~Mj\"ornmark$^g$,
P.~Conde Mu\'i\~no$^e$,
P.~Puzo$^b$,
M.~Rijssenbeek$^k$,
J.G.~Saraiva$^{d,e}$,
N.~Seguin-Moreau$^b$,
S.~Soares$^{d,e}$,
H.~Stenzel$^h$,
M.~Thioye$^k$,
D.~Varouchas$^b$~
and V.~Vorobel$^j$\\
\llap{$^a$}CERN, PH Department, Geneva, Switzerland,\\
\llap{$^b$}Laboratoire de l'Accelerateur Lineaire, Orsay, France\\
\llap{$^c$}CEFITEC/Faculdade de Ci\^encias e Tecnlologia da Universidade Nova de Lisboa, Lisbon, Portugal\\
\llap{$^d$}CFNUL/Faculdade de Ci\^encias da Universidade de Lisboa, Lisbon, Portugal\\
\llap{$^e$}LIP -- Laborat\'orio de Fis\'ica Experimental e Instrumenta\,c\~ao em Part\'iculas, Lisbon, Portugal 
\llap{$^f$}Ecole Polytechnique, Palaiseau, France\\
\llap{$^g$}University of Lund, Lund, Sweden\\
\llap{$^h$}II. Physikalisches Institut, Justus-Liebig-Universit\"at, Giessen, Germany\\
\llap{$^i$}IFJ PAN, Krakow, Poland\\
\llap{$^j$}Faculty of Mathematics and Physics, Charles University in Prague, Czech Republic\\
\llap{$^k$}Stony Brook University, New York, USA\\
E-mail: \email{Stefan.Ask@cern.ch}
} 

\abstract{
A scintillating fibre tracker is proposed to measure elastic proton scattering at very small angles in the ATLAS experiment at CERN. 
The tracker will be located in so-called Roman Pot units at a distance of $240~{\rm m}$ on each side of the ATLAS interaction point.
An initial validation of the design choices was achieved in a beam test at DESY in a relatively low energy electron beam and using 
slow off-the-shelf electronics.
Here we report on the results from a second beam test experiment carried out at CERN, where new detector prototypes were tested in a 
high energy hadron beam, using the first version of the custom designed front-end electronics. The results show an adequate tracking 
performance under conditions which are similar to the situation at the LHC. In addition, the alignment method using so-called overlap 
detectors was studied and shown to have the expected precision. 
}


\begin{document}

\section{Introduction}

New physics at the LHC might manifest itself as deviations from the Standard Model predictions of cross sections and thus the 
normalization of the measured cross sections is of utmost  importance. The uncertainty of the absolute luminosity is the main factor 
limiting the measurements precision. Traditionally, the absolute luminosity at hadron colliders was determined via elastic scattering 
of protons at small angles. This is also one of the approaches pursued by the ATLAS experiment at the LHC. ATLAS is aiming to measure 
at such small angles that the elastic scattering becomes sensitive to the well-known electro-magnetic amplitude.
A scintillating fibre tracker system, called \alfa , is used to measure the scattered protons. Fibre tracker modules will be inserted 
in so-called Roman Pot units which makes it possible to detect scattered protons very close to the beam. The vertically arranged Roman 
Pots penetrate into the LHC beam pipe on each side of the ATLAS experiment at a distance of $240~{\rm m}$ from the interaction point. 
The luminosity measurement and the requirements on the detector have been described in \cite{bib:TDR}. The main requirements which are 
directly related to the fibre tracker design, such as light yield, sensitivity at the fibre edge etc., were shown to be fulfilled in a 
beam test of \alfa\ at DESY in November 2005 \cite{bib:NIM}.

A major difference between the 2005 setup and the final system is the dedicated \alfa\ front-end (FE) electronics which will be used in 
ATLAS. The previously used analog readout that was based on standard (NIM and VME) electronics does not meet requirements at the LHC, 
such as a $40~{\rm MHz}$ clock frequency and intermediate data storage during the trigger latency time. For this reason, dedicated FE 
electronics have been designed for \alfa . The new electronics are entirely based on binary hit information. The number of channels, the 
available space and the remote location of the final detectors are challenges for the design of the FE electronics. To operate \alfa\ 
with the final components is therefore a crucial validation step of the overall system design.

The measured spatial resolution in the previous test of \alfa\ was shown to be limited by multiple scattering of the relatively low 
energy, $6~{\rm GeV/c}$, electron beam. This also motivated a second test of the fibre tracker in a high energy hadron beam, which 
approaches the situation in the LHC and where effects due to multiple scattering are negligible. 
The results from the DESY test together with GEANT4 simulations made it possible to predict that the obtained spatial resolution of 
$36~{\rm \mu m}$ would decrease to about $20-25~{\rm \mu m}$ if \alfa\ was operated in a high energy hadron beam.

The precise relative vertical distance of the upper and lower \alfa\ detectors in one Roman Pot unit cannot be derived from the physics 
data, but will be measured by the so-called Overlap Detectors (OD), also based on scintillating fibre technology. The ODs need to measure 
only the vertical coordinate. In the final \alfa\ setup the ODs will be located in special extrusions of the Roman Pots which will 
gradually overlap when the upper and lower pots are moved towards the beam axis. These detectors, as well as the alignment method, were 
tested for the first time in this experiment.

\section{The CERN testbeam setup}

The beam test was carried out at the CERN SPS H8 beam which is a secondary beam in the CERN North Area. The primary target consists of 
a $30~{\rm cm}$ long Beryllium plate. We have used for most of the period a secondary beam at $230~{\rm GeV/c}$ (70\% $p$ and 30\% 
$\pi^+$) and for a few days a $20~{\rm GeV/c}$ secondary beam (mainly $\pi^{+}$ beam with a small $p$ contamination). The beam spot size 
was between $1-10~{\rm mm}$, and varied with the beam line settings, and the beam divergence was about $0.15~{\rm mrad}$.

The \alfa\ detector was mounted on a table which allowed the detector to be translated in the direction of the axes of the plane 
transverse to the beam, as well as being rotated around these axes. The trigger was based on a $30 \times 30 ~{\rm mm^2}$ plastic 
scintillator and its signal was put in coincidence with the signal from a second scintillator that had dimensions similar to the active 
area of the particular \alfa\ prototype used.

\subsection{Prototype fibre detectors}

\begin{figure}
\begin{center}
 \includegraphics[width=.49\textwidth]{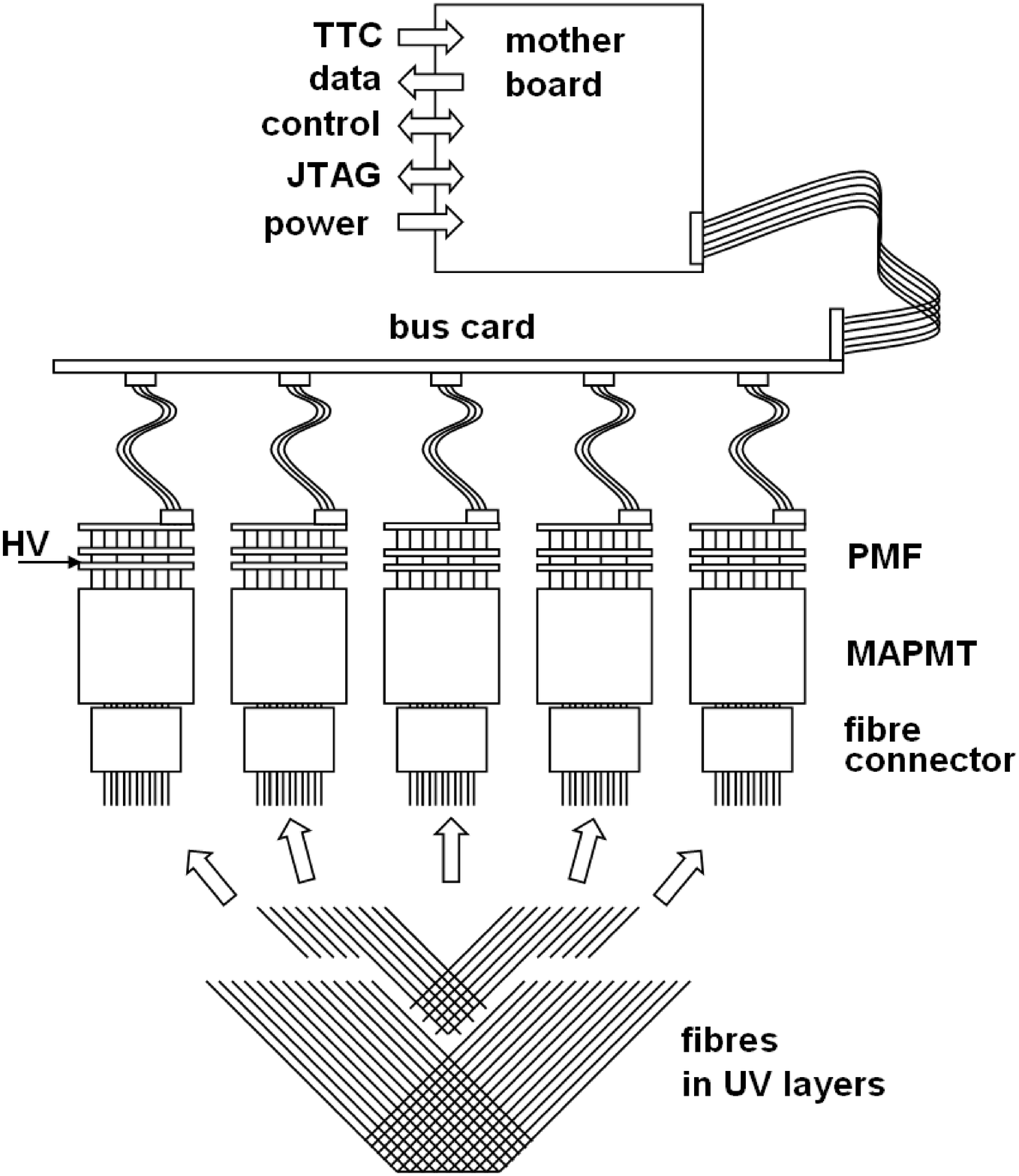} 
 \includegraphics[width=.49\textwidth]{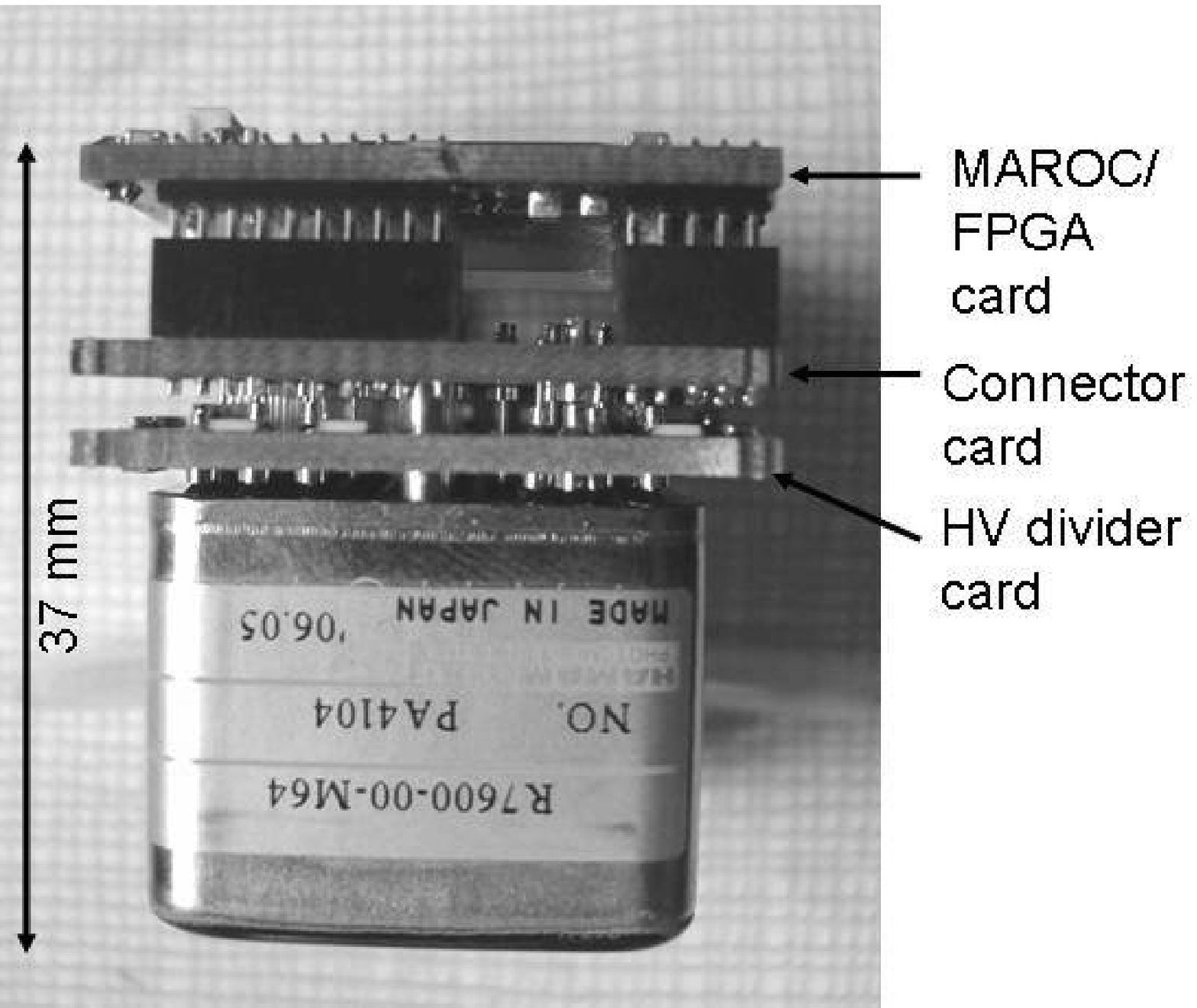}
\caption{Schematic representation of the testbeam system setup (left). The MAPMT together with the three PCBs making a PMF unit (right).}
\label{fig:tb_setup}
\end{center}
\end{figure}

Two new \alfa\ prototypes \cite{bib:METROLOGY, bib:METROLOGY2} were built with an increased number of fibres compared to the DESY 
testbeam, however, their design and construction followed the same principles. A detector plane consists of a ceramic substrate which 
on both sides supports Aluminium coated square fibres\footnote{ Fibre type SCSF-78, single cladded, S-type, from Kuraray, Japan} of 
$500~{\rm \mu m}$ thickness in a so-called UV configuration, i.e. the two fibre layers have an angle of $\pm 45^\circ$ relative to the 
vertical axis as shown in the left part of figure \ref{fig:tb_setup}. In the final detector, ten such UV planes with two times 64 
fibres will be assembled with a relative staggering of multiples of $\sqrt{2} \cdot 50~{\rm \mu m}$.

The first detector constructed for the testbeam, called \alfaA , included two planes which contained the full set of 64 fibres per 
layer. It was tested with a prototype of the final trigger scintillator whose shape matches the overlap area of the fibres. It was 
connected via a specially formed light guide to a PMT\footnote{Hamamatsu H3164. Active diameter 8 mm.}. The challenges with this 
trigger counter is the stringent space limitations in the pot and the requirement of a very uniform trigger efficiency in order not 
to bias the measurement. Lab tests with a Sr-90 source showed a photoelectric yield in excess of 30. Setting the trigger threshold 
at $\approx 5~{\rm p.e.}$ should guarantee an excellent uniformity. The second \alfa\ tracker comprised of 10 staggered planes with 
16 fibres per layer (\alfaB) and was used for resolution studies.

In addition to the two \alfa\ trackers, two prototype Overlap Detectors were built \cite{bib:OD1}. They consist of horizontally 
mounted scintillating fibres of the same type and size as the main detector. The fibres are organized in two staggered layers with 
30 fibres each. The two detectors were placed in the beam such that they overlapped vertically. Their relative vertical position 
could be mechanically controlled by a micrometric screw with a precision of a few microns.

The scintillating fibres were routed over typically $25~{\rm cm}$ to custom designed fibre connectors with 64 channels in a 
$8 \times 8$ configuration, which matches precisely the segmentation of the multi anode PMTs\footnote{MAPMT type R7600, Hamamatsu, 
Japan}. The connectors fit in slots of the Roman Pot vacuum flange, on the top (air) side of which are mounted the MAPMTs with their 
front-end electronics.

All detectors underwent careful metrological analysis with a 3D coordinate measurement machine. The positions and slopes of all 
fibres were individually determined and stored in a database. These parameters were then used during the offline-analysis for the 
space point reconstruction.

\subsection{The \alfa\ electronics and system setup}

A very compact design was adopted for the dedicated \alfa\ FE electronics, where a large part of the electronics is directly mounted 
onto the MAPMTs. 
The right part of figure \ref{fig:tb_setup} shows the so-called PMF (PhotoMultiplier Front-end) unit, consisting of three PCBs mounted 
on top of the MAPMT. From the bottom, the first board is a high voltage (HV) divider which is followed by a board which routes the 
relevant signals from the MAPMT to the third board. The third board contains the multi anode readout chip (MAROC) \cite{bib:MAROC}, 
providing amplifiers and discriminators for each of the 64 MAPMT channels, and a FPGA for handling the readout. 

The readout of the detector follows the ATLAS requirements \cite{bib:L1,bib:DAQ} and is controlled by a motherboard. When the trigger 
signal arrives to the \alfa\ detector it is received and fanned-out by the motherboard to the different PMFs. Each PMF stores the 
data from the MAPMT at $40~{\rm MHz}$ in a pipeline and at the arrival of a trigger it sends back the data from the corresponding 
pipeline location. Due to the synchronous trigger and readout system, the latency of the trigger has to be timed in carefully with 
respect to the data in the pipeline. The data from all the PMFs is then collected by the motherboard where it is formated and sent 
through a gigabit optical link (GOL) \cite{bib:GOL} to the readout system. The configuration of the FE components is done using an 
embedded local monitor board (ELMB) \cite{bib:DAQ} which is part of the motherboard.
 
The first version of the \alfa\ FE electronics was used during the testbeam period and the setup consisted of 5 PMFs, distributed 
in a row (as is foreseen for the final detector), and a motherboard. The left part of figure \ref{fig:tb_setup} shows a schematic 
view of the setup, with the motherboard at the top and the cabling structure to distribute signals to the 5 PMFs. It should be noted 
that the interconnection between the motherboard and the PMFs will be based on light Kapton cables in the final system, to replace 
the large flat cables and adapters used at the testbeam. 

A computer with PVSS was used to configure the electronics installation in the testbeam area, through a CAN BUS and the ELMB. The 
trigger signal was sent through an optical fibre to the motherboard by standard components developed for the ATLAS timing, trigger 
and control (TTC) \cite{bib:L1} system.
The trigger-busy logic was made using standard NIM electronics and the DAQ software was implemented within the standard ATLAS frame 
work. A readout PC was equipped with a FILAR card \cite{bib:FILAR} to read data through the GOL that was connected to the motherboard. 
The event rate was restricted by the ad hoc implementation of the readout system where a peak rate of about $1.5~{\rm kHz}$ and an 
average rate of about $250~{\rm Hz}$ were obtained.

In the CERN testbeam all the components of the final system were used except for the so-called ROD (ReadOut Driver) card which forms 
the interface between the GOLs and the common ATLAS readout system. During the tests of the electronics several imperfections were 
identified. These were strongly suspected of creating fake noise hits (as discussed below) and will be corrected in the next version 
of the electronics.

\section{The data analysis}

After converting the raw data into an appropriate format, the analysis was entirely based on the ROOT analysis software \cite{bib:ROOT}. 
A set of basic analysis routines were used during the testbeam to monitor the data during the data-taking, however, the main analysis 
was carried out after the testbeam period.

After an initial running period, where the trigger and readout system was timed in and the basic detector operation was established, 
the beam time was split into one period for each detector prototype. Between $2-7$ million events were recorded with each detector 
and the high energy proton beam was used for the runs dedicated to resolution studies. In addition, some 100k events were recorded 
together with the setup of the DEPFET collaboration who were starting tests after the ALFA period with a high precision Si-telescope 
in the same zone.

\subsection{The \alfa\ data quality}

The overall quality of the data taken during the beam period was fully satisfactory. No problems due to bad synchronization or data 
corruption were encountered. However, as mentioned before, there were several imperfections in the electronics that were suspected 
of creating fake noise hits. In order to disentangle the possible different noise sources (fibres, fibre-PMT interface, electronics) 
we exploited different experimental configurations:
\begin{itemize} 
\item[A)] A complete MAPMT or individual MAPMT channels, that were not connected to fibres (e.g. in \alfaA\ or \OD );
\item[B)] MAPMTs that were connected to, \OD , fibres that were not in the beam;
\item[C)] MAPMTs that were connected to fibres that were in the beam.
\end{itemize}

The observed probability of registering a hit was calculated as the number of measured hits in the whole MAPMT divided by the number 
of channels (64) and the number of events. For our standard HV and threshold setting, the overall noise level for MAPMTs without any 
fibres connected (case A) was of the order of 0.02\%, increasing to 0.4\% for MAPMTs that were connected to fibres, when the fibres 
were not in the beam (case B). 

The noise level also showed correlations with the number of hits registered in the MAPMT which hints at some coupling in the read out 
chain (analogue or digital). 

For MAPMTs which read out fibres that were in the beam (case C), the noise was studied by the occurrence of neighbours of a selected 
hit. For a given hit, the probability of registering a hit in a direct neighbouring MAPMT channel was of the order of 10\%. This level 
decreased to about 5\% when only hits associated to reconstructed tracks were considered. This selection eliminates randomly distributed 
noise hits which were typically produced in groups with several hits at the same time. The existence of correlated noise was also 
supported by the analysis of the hit multiplicity per fibre layer. Ideally it should be one (assuming 100\% efficiency, zero cross-talk 
and no noise), however the measured hit multiplicity showed a mean of two with tails of several tens of fibres hits.    

\subsection{Space point reconstruction}

One of the main goals of this test experiment was to measure the intrinsic spatial resolution of the detector. As mentioned earlier 
the value found at the DESY testbeam ($\sigma_x = \sigma_y = 36~{\rm \mu m}$), was limited mainly by multiple scattering in the 
relatively low energy electron beam and should significantly improve in a high energy hadron beam. 

All studies of the tracking performance were based on the data taken with the \alfaB\ detector.
The problems with the data quality, discussed above, called for an improved tracking algorithm, which was largely immune to fake noise 
hits. In a first stage, hit candidates were selected based on a seed track obtained by a Hough transform method \cite{bib:HOUGH}. Here 
the closest hit to the seed track in each \alfa\ layer was selected under the condition that it was within a distance of 2 fibre widths 
from the seed track. The selected hits were then used by a so-called minimum overlap algorithm \cite{bib:NIM} in order to reconstruct 
the impact point in the $x-y$ plane transverse to the beam. This algorithm assumes that the tracks are perpendicular to the \alfa\ 
layers and the $z$ coordinate information therefore becomes irrelevant.

\subsubsection{Stand-alone method}

In the first part of the resolution study, tracks were reconstructed by \alfa\ as two independent track segments, using planes 1-5 
(layers 1-10 = H1) and planes 6-10 (layers 11-20 = H2) separately. In this way the \alfa\ resolution can be studied without an external 
tracking detector, however, with limited possibilities to estimate the efficiency. In the following we call this method the stand-alone 
method. For this study, events were only used if their $x-y$ position was within a $2~{\rm mm}$ radius from the centre of \alfa\, which 
ensured that the selected particles traversed a region of \alfa\ where all 20 layers physically overlap. All following studies were 
done for both the $x$ and $y$ coordinates, but for simplicity only one will be discussed if the results for the other coordinate were 
the same.

\begin{figure}
\begin{center}
 \includegraphics[width=.49\textwidth]{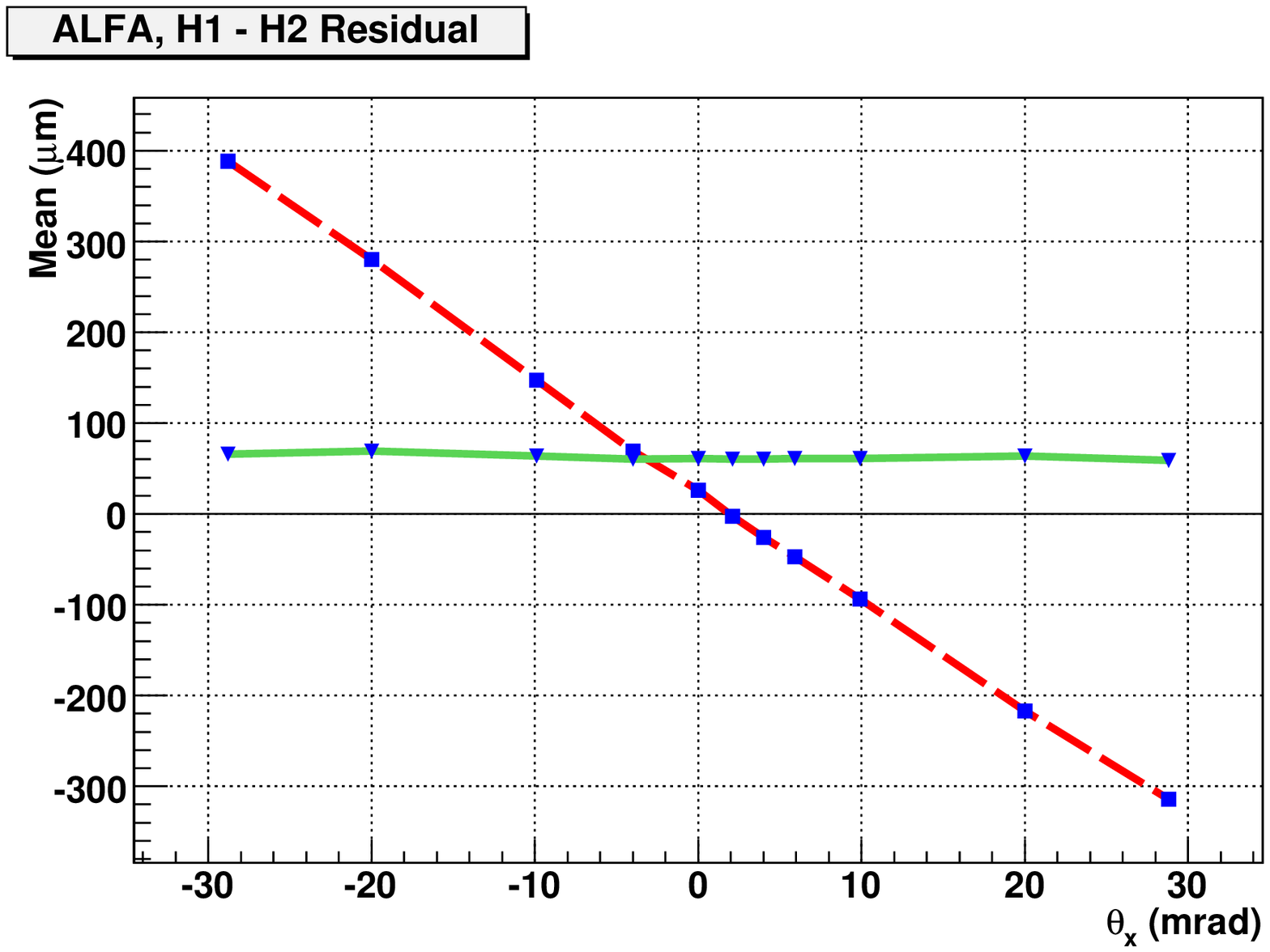} 
 \includegraphics[width=.49\textwidth]{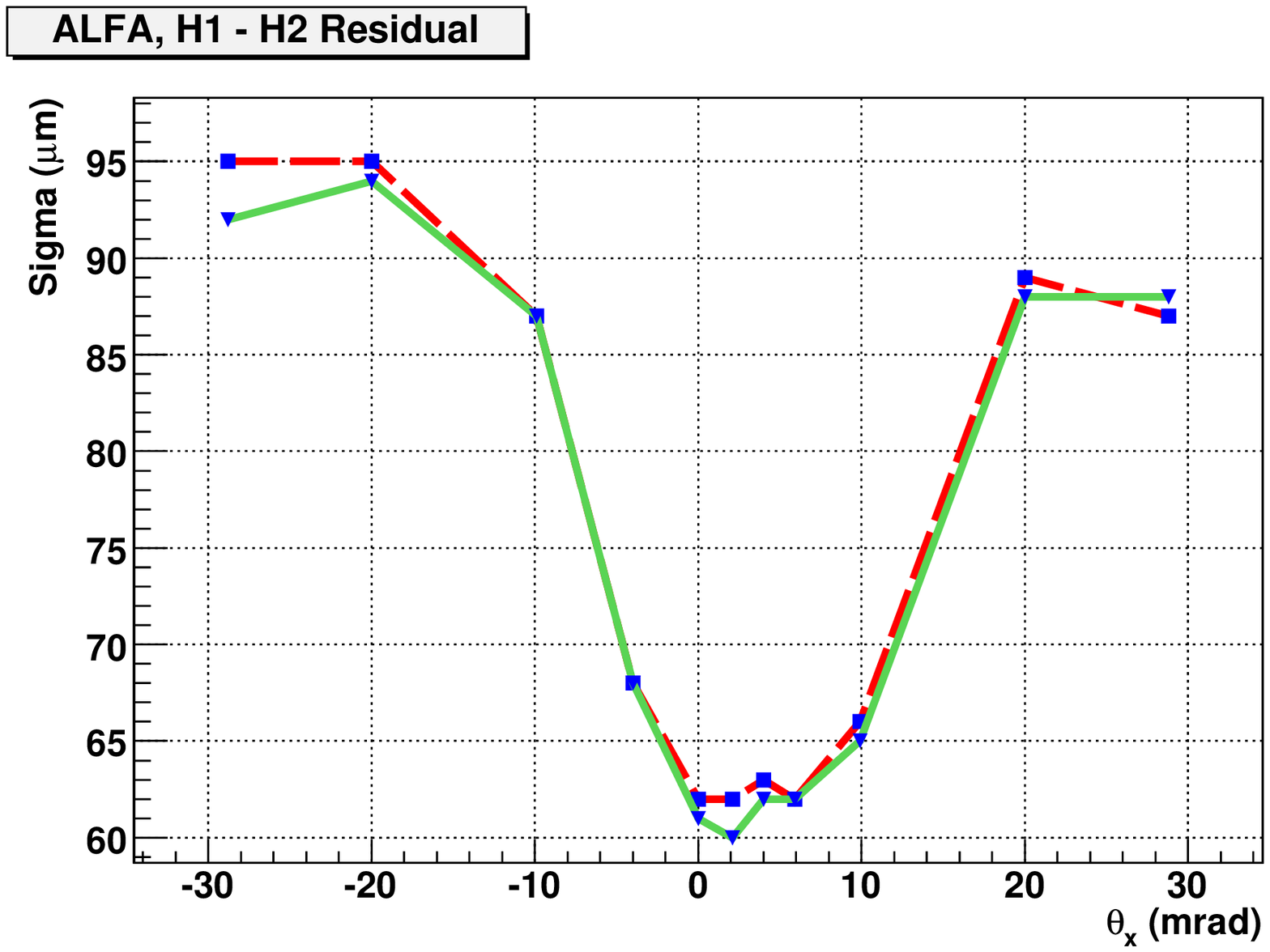}
\caption{ The mean and sigma of the $x_{H1}-x_{H2}$ distribution as a function of the $\theta_x$ angle.
The solid line is based on the reconstructed $x$ coordinates and the dashed line on the $y$ coordinates.}
\label{fig:mean_sigma_tx}
\end{center}
\end{figure}

Before taking data dedicated for resolution studies, an angular scan was made in order to align \alfa\ with respect to the beam axis. 
Runs were taken with \alfa\ rotated around both the $x$ and $y$ axis and during these runs the MAPMTs were operated at $900~{\rm V}$. 
The runs were analyzed based on the mean value and sigma of the $x_{H1}-x_{H2}$ track residual distributions. 
Figure \ref{fig:mean_sigma_tx} shows the mean (left) and sigma (right) of the residual distribution as functions of the rotation angle 
around the $x$-axis, $\theta _x$, with respect to the original position of \alfa . 

Since the minimum overlap algorithm is reconstructing the $x$ and $y$ coordinate of the particle separately for the two halves, the 
presence of an angle introduces a systematic shift of the mean of the distribution. This is clearly seen in figure \ref{fig:mean_sigma_tx} 
where the mean in $y$ shifts linearly with the $\theta_x$ angle and where a perfectly perpendicular alignment of \alfa\ to the beam 
corresponds to a mean equal to zero. The shift of the mean therefore provides a method for a precise alignment ($\sim 1 ~{\rm mrad}$), 
which will also be useful when operating at the LHC. Figure \ref{fig:mean_sigma_tx} also shows the residual sigma as a function of 
$\theta_x$ where the minimum is consistent with an optimal alignment of about $\theta_x = 2 ~{\rm mrad}$ obtained from the results 
based on the mean of the distribution.

\begin{figure} 
\begin{center}
 \includegraphics[width=.6\textwidth]{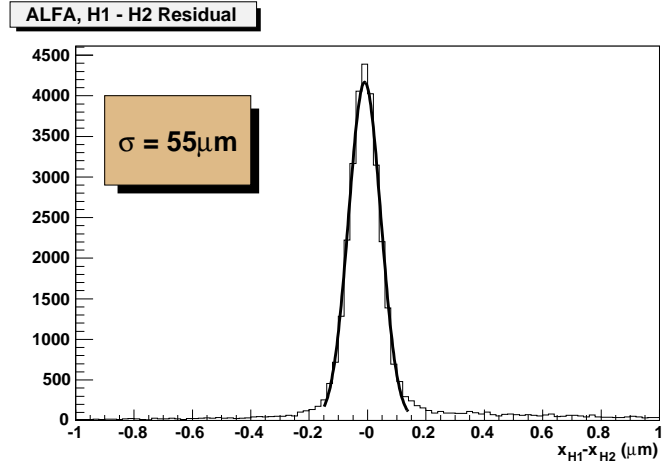}
\caption{ The residual distribution from the $x$-coordinates reconstructed by the two \alfa\ halves, $H1$ and $H2$.}
\label{fig:x_sa_res}
\end{center}
\end{figure}
After the alignment, dedicated high statistics runs were taken in order to study the spatial resolution of \alfa. During these runs 
\alfa\ was operated both at $900~{\rm V}$ and $950~{\rm V}$. Figure \ref{fig:x_sa_res} shows the residual distribution obtained from 
high statistics runs taken at $950~{\rm V}$. A Gaussian fit to the distribution gives $\sigma = 55 \pm 0.3 ~{\rm \mu m}$ (only 
statistical error from the fit) and as both half detectors are practically identical, the spatial resolution of each of them is 
expected to be $\sigma_{H1} = \sigma_{H2} = 55 / \sqrt{2}\sim 39 ~{\rm \mu m}$. 

Comparison with the output from a geometrical Monte-Carlo (MC) \cite{bib:METROLOGY}, with the as-measured detector geometry as input, 
allows us to interpret the measured value as a spatial resolution of the full \alfa\ detector,
\begin{equation}
\sigma_{ALFA} = \frac{\sigma_{H} \ominus \sigma_{geo}}{2} \oplus \sigma_{geo} = \frac{\sqrt{\sigma_{H}^2 + 3 \cdot \sigma_{geo}^2}}{2} 
\label{eq:ALFA_RES}
\end{equation}
where $\sigma_{H}$ is the measured half \alfa\ resolution and $\sigma_{geo}$ is the constant contribution from geometrical imperfections 
of the detector estimated by the MC to be $14~{\rm \mu m}$. Based on this formula the measured half resolution of $39~{\rm \mu m}$ 
indicates a full detector resolution of about $23~{\rm \mu m}$. Several systematic effects were studied, e.g. alternative methods to 
obtain $\sigma_{geo}$ and different ranges for the Gaussian fit, and from these a total error of $\pm 2~{\rm \mu m}$ was obtained. 

\subsubsection{\alfa\ versus Si-telescope}

A dedicated reference tracking detector was not available during the regular \alfa\ testbeam period. A few runs could, however, be 
taken with track information from a high precision telescope, situated about $1~{\rm m}$ upstream of the \alfa\ setup, with which 
the DEPFET collaboration performed measurements. The data acquisition systems of the two experiments were synchronized during these 
runs. 
The tracks reconstructed by the telescope were extrapolated to the $z$-position of \alfa\ and after a software alignment of the 
telescope with the \alfa\ coordinate system, the track positions of both systems could be directly compared. Two scenarios were 
considered. In the first, the reconstruction efficiency was maximized by using loose cuts. In the second scenario, tighter cuts were 
applied on the number of layers with more than one hit and on the total amount of hits. With these quality cuts the resolution was 
optimized. 

\begin{table}[h!]
\begin{center}
\begin{tabular}{|c|c|c|}\hline
& reconstruction cuts & quality cuts \\ \hline
$\sigma_H$ [${\rm \mu m}$] & $56 \pm 3$ & $51 \pm 2$ \\ \hline
$\sigma_T$ [${\rm \mu m}$] & \multicolumn{2}{|c|}{46$\pm$2 in $x$ and 66-77 $\pm$4 in $y$} \\ \hline
$\sigma_{ALFA}$ [${\rm \mu m}$] & $26 \pm 3$ & $25 \pm 3$ \\ \hline
Efficiency [$\%$] & $96 \pm 3$ & $56 \pm 3$ \\ \hline
\end{tabular}
\caption[Resolution and efficiency]{\label{tab:reseff}{Resolution (averaged over $x$ and $y$) and efficiency.}}
\end{center}
\end{table}
The residuals from the half-detectors and the telescope were studied in a first stage and from knowing $\sigma_{H}$, the resolution 
of the telescope, $\sigma_T$, at the position of \alfa\ was determined. Due to the $1~{\rm m}$ distance between the telescope and 
\alfa , the intrinsic resolution of about $10 ~{\rm \mu m}$ inside the telescope was degraded to an observed telescope resolution 
at the \alfa\ position of about $46~{\rm \mu m}$ in $x$ and to $66-77 ~{\rm \mu m}$ in $y$. The performance of the telescope in the 
vertical direction varies with the hit region, which explains the observed variation of the resolution. 

In a second stage, the spatial resolution of the full \alfa\ detector could be extracted. This was done by subtracting 
the calculated telescope contribution from the \alfa\ -- telescope residual distribution. At the same time the \alfa\ efficiency was 
determined. The results are summarized in table~\ref{tab:reseff} where it is shown that with basic reconstruction cuts a detection 
efficiency above $95\%$ was obtained with a spatial resolution of about $26~{\rm \mu m}$. Both the stand-alone study and the study 
using the telescope showed that the resolution improves marginally for the full detector when quality cuts were applied while the 
efficiency drops drastically. The modest telescope resolution at the position of \alfa\ and the limited statistics implies relatively 
large errors, as seen in table~\ref{tab:reseff}, which were obtained from the spread of results in analyses using different runs. 

\subsection{The overlap detectors}

\begin{figure}
\begin{center}
 \includegraphics[width=.6\textwidth]{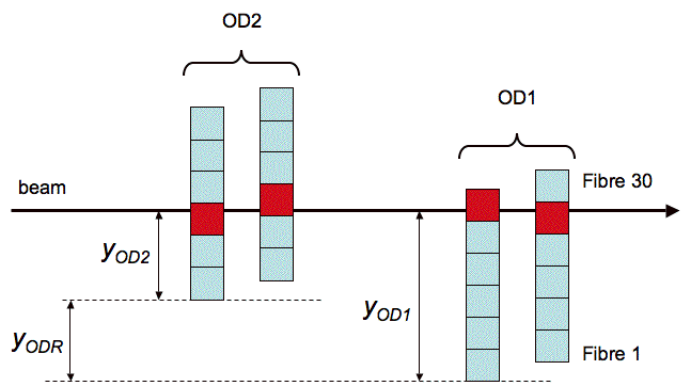}
\caption{ The vertical distance between the two overlap detectors is calculated from the measurement of particles which traverse both 
detectors: $y_{ODR}=y_{OD2}-y_{OD1}$.}
\label{fig:od1}
\end{center}
\end{figure}
\begin{figure}
\begin{center}
 \includegraphics[width=.49\textwidth]{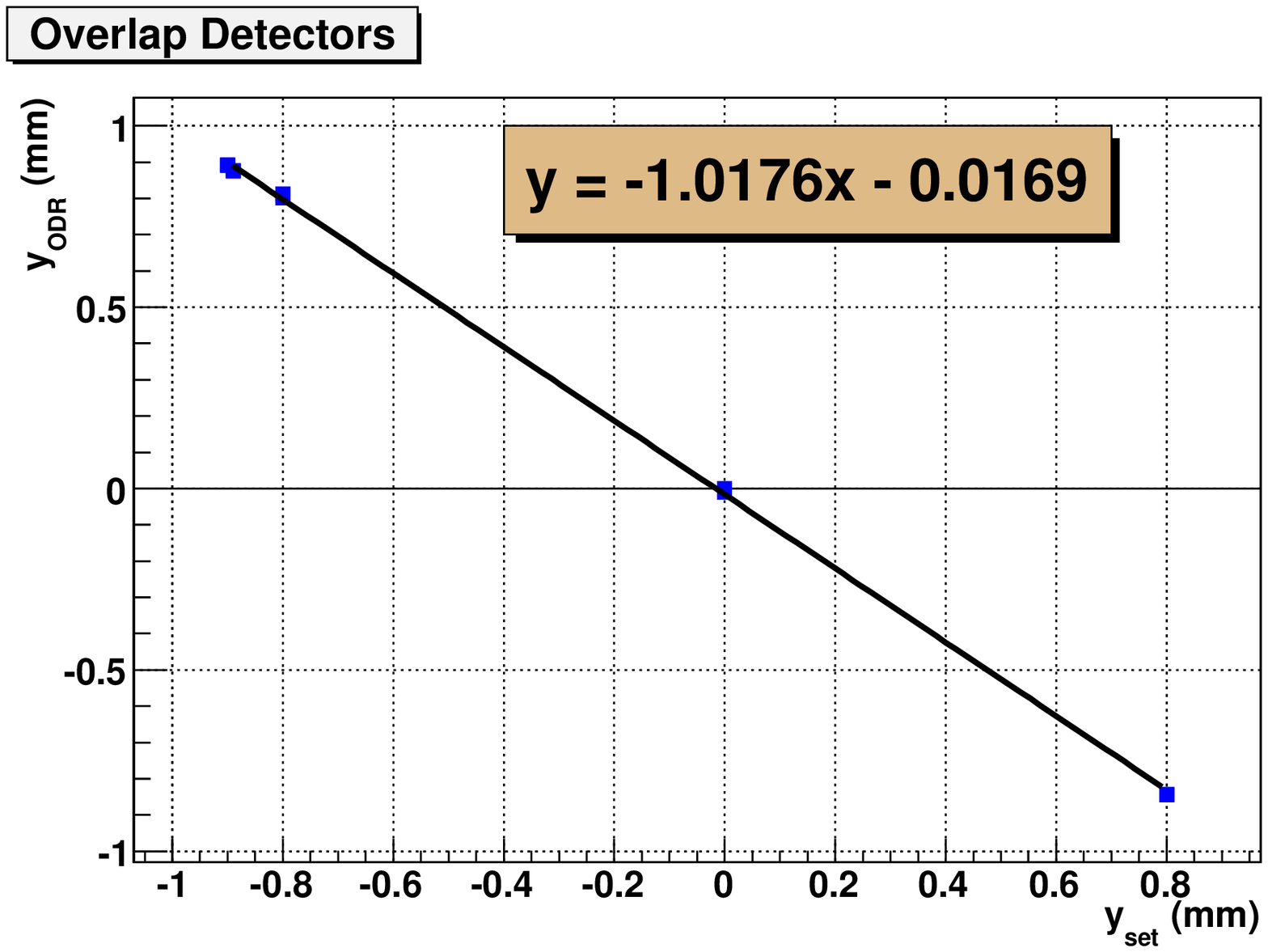}
 \includegraphics[width=.49\textwidth]{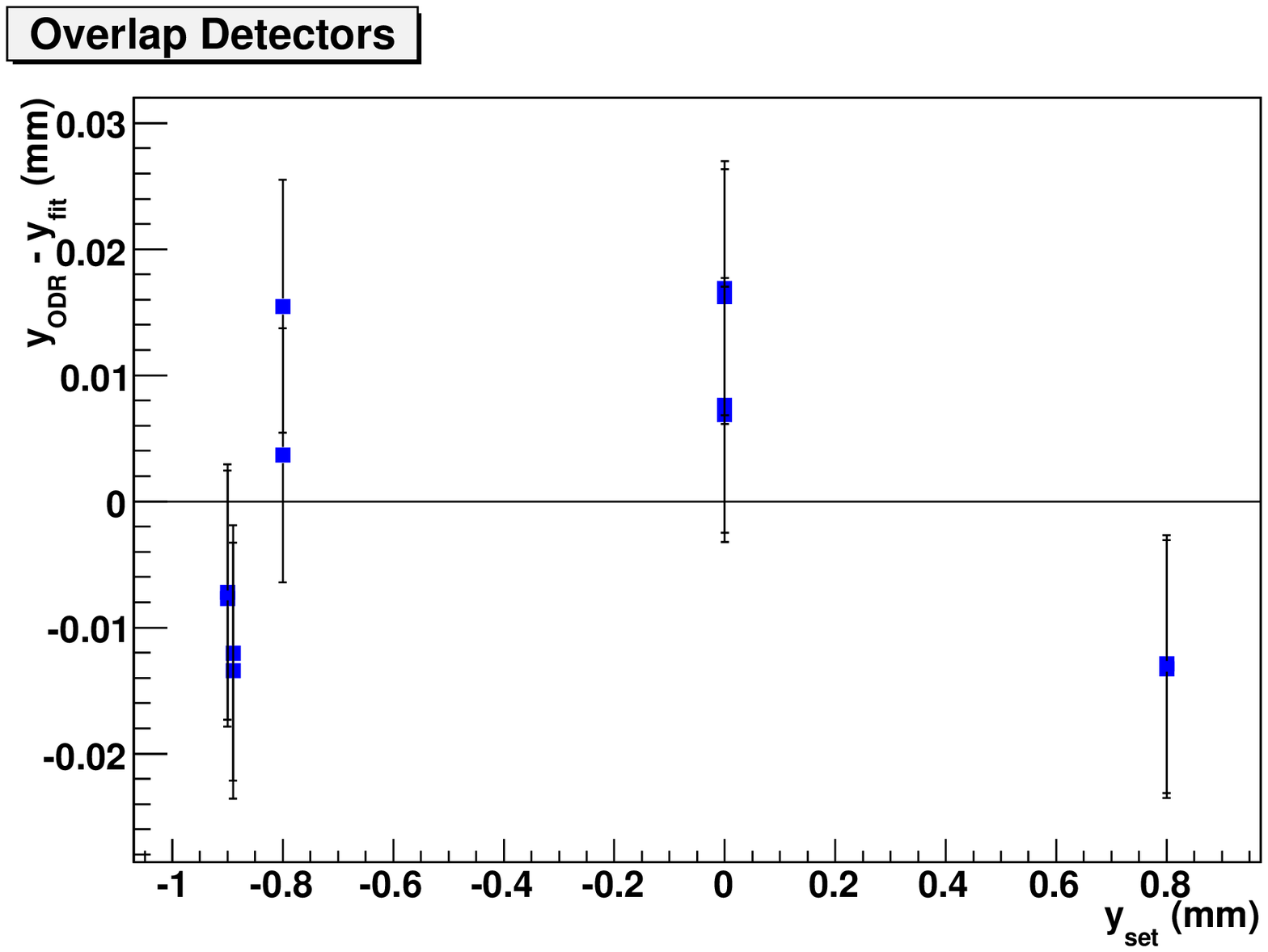}
\caption{ Reconstructed vertical distance $y_{ODR}$ between the two overlap detectors as a function of the mechanically set distance 
$y_{set}$ (left). The straight line is a fit to the data points. Difference between the reconstructed distance $y_{ODR}$ and the fit 
for the same data (right). The error bars represent the quadratic sum of the errors from the reconstruction algorithm and the errors 
of the mechanical measurement.}
\label{fig:od2}
\end{center}
\end{figure}

The vertical distance between the two overlap detectors was reconstructed as the average difference between the 
positions of the fibres that were hit in each detector
\begin{equation}
y_{ODR} = \frac{1}{n}\sum_{i=1}^n  \left[ \frac{1}{2} ( y_{OD2,1} + y_{OD2,2} ) - \frac{1}{2} ( y_{OD1,1} + y_{OD1,2} ) \right] 
\end{equation}
where $y_{ODk,l}$ is the vertical position of the fibre hit in the layer $l$ of the detector $k$ and the sum is over the number of 
selected events (see figure \ref{fig:od1}). The events were selected on the basis of the number of fibres that were hit. Only events 
with a total of 4 hits and a single hit per plane were used for the reconstruction. The efficiency was only 11\% due to the noise 
problems mentioned above. A good correlation was found between the reconstructed relative distance $y_{ODR}$ and the set value 
$y_{set}$ as shown in the left part of figure \ref{fig:od2}. The difference between $y_{ODR}$ and the straight line fit is in general 
smaller than $15 ~{\rm \mu m}$ (see right part of figure \ref{fig:od2}). Most of the positions were independently measured a second 
time. The two sets of reconstructed positions are compatible within the precision of the micrometric screw.

\section{Conclusions}

The ATLAS experiment is planning to measure the absolute luminosity from elastic scattering at very small angles using a scintillating 
fiber tracker called \alfa . The \alfa\ fiber detector was validated in a testbeam at DESY in November 2005, however, with standard 
electronics and using a $6~{\rm GeV/c}$ electron beam.

A second testbeam was carried out at CERN using new prototypes together with the first version of the dedicated \alfa\ electronics, 
required for operation within ATLAS at the LHC, and using a high energy hadron beam. The high energy hadron beam approaches the final 
situation in the LHC and the spatial resolution was expected to improve with respect to the previous test because of reduced multiple 
scattering. A number of imperfections, likely to create fake noise hits, were identified during the tests of the electronics which 
will be corrected in the next version. An increased noise rate was also observed in the data with respect to the previous test where 
standard electronics was used, however, the tracking performance was still shown to be adequate for the purpose of the luminosity 
measurement. The space point reconstruction was studied by two different methods which showed resolutions of about $25 ~{\rm \mu m}$ 
and where a reconstruction efficiency above $95 \%$ was obtained.

Overlap detectors will be used for a precise vertical alignment of \alfa\ at the LHC. These detectors, as well as the alignment method, 
were tested for the first time in the CERN testbeam and it was shown that the required precision could be achieved, since the vertical 
relative alignment could be controlled to a precision better than $15 ~{\rm \mu m}$.

The natural next step will be to construct a full-size detector consisting of 10 planes with 2 x 64 fibres. This full-size detector 
will be integrated in one Roman Pot unit and together with the overlap detectors and the trigger counters make a full system. We 
would also include all the  front-end electronics together with the mother board. Such a complete system will allow us to evaluate 
and eliminate possible remaining electronics noise sources. Moreover the integration in the Roman Pot  will permit  us to address 
mounting and alignment precision in the Roman Pot and also relative alignment precision between the main detectors and the overlap 
detectors.  
 
\section{Acknowledgments}
We would like to thank our technical staff for their competent work in building the fibre detectors and preparing the testbeam setup: 
A. Folley, L. Kottelat, J. Mulon, M. v. Stenis (all CERN), M. Szauter (University of Giessen) and J. Patriarca (LIP). We are grateful 
to the ALICE Silicon Pixel team, M. Burns and P. Riedler, for providing us with a motorized XYR table. We wish to acknowledge the work 
of the ATLAS TDAQ community in providing the underlying online software and infrastructure for read-out and dataflow. We would like 
to express our gratitude to M. Mathes and the DEPFET collaboration for allowing us to take common runs with their Silicon beam 
telescope and for providing us with the tracking data. We would like to acknowledge the effort of the SPS team for delivering stable 
beams and their help with the beam setup and tuning. Acknowledgments also to FCT/MCTES for grants and funding of the project 
"Collaboration in the ATLAS Experiment" POCI/FP/63936/2005.



\end{document}